\documentclass[preprint,12pt]{elsarticle}%
\usepackage{amssymb}
\usepackage{amsfonts}
\usepackage{amsmath}
\usepackage{graphicx}%
\providecommand{\U}[1]{\protect\rule{.1in}{.1in}}
\graphicspath{{G:/ANTONIO/dottorato_(HD)/Dimple/Testo/figure/}}
\journal{journal}

\begin{document}
%
%TCIMACRO{\TeXButton{Begin frontmatter}{\begin{frontmatter}}}%
%BeginExpansion
\begin{frontmatter}%
%EndExpansion



%TCIMACRO{\TeXButton{Title}{\title
%{Universal features in "stickiness" criteria for soft adhesion with rough surfaces}%
%}}%
%BeginExpansion
\title
{Universal features in "stickiness" criteria for soft adhesion with rough surfaces}%
%EndExpansion


%TCIMACRO{\TeXButton{Author}{\author{M. Ciavarella}}}%
%BeginExpansion
\author{M. Ciavarella}%
%EndExpansion
%

%TCIMACRO{\TeXButton{Address}{\address
%{Politecnico di BARI. DMMM dept. V Japigia 182, 70126 Bari. }
%\address{email: mciava@poliba.it }}}%
%BeginExpansion
\address{Politecnico di BARI. DMMM dept. V Japigia 182, 70126 Bari. }
\address{email: mciava@poliba.it }%
%EndExpansion
%

%TCIMACRO{\TeXButton{Begin abstract}{\begin{abstract}}}%
%BeginExpansion
\begin{abstract}%
%EndExpansion

A very interesting recent paper by Dalvi \textit{et al.} has demonstrated
convincingly with adhesion experiments of a soft material with a hard rough
material that the simple energy idea of Persson and Tosatti works reasonably
well, namely the reduction in apparent work of adhesion is equal to the energy
required to achieve conformal contact. We demonstrate here that, in terms of a
stickiness criterion, this is extremely close to a criterion we derive from
BAM (Bearing Area Model) of Ciavarella, and not very far from that of Violano
\textit{et al.} It is rather surprising that all these criteria give very
close results and this also confirms stickiness to be mainly dependent on
macroscopic quantities.%

%TCIMACRO{\TeXButton{End abstract}{\end{abstract}}}%
%BeginExpansion
\end{abstract}%
%EndExpansion
%

%TCIMACRO{\TeXButton{Begin keyword(s)}{\begin{keyword}}}%
%BeginExpansion
\begin{keyword}%
%EndExpansion

Adhesion, JKR model, DMT\ model, soft matter, roughness models.%

%TCIMACRO{\TeXButton{End keyword(s)}{\end{keyword}}}%
%BeginExpansion
\end{keyword}%
%EndExpansion
%

%TCIMACRO{\TeXButton{End frontmatter}{\end{frontmatter}}}%
%BeginExpansion
\end{frontmatter}%
%EndExpansion



\section{Introduction}

\begin{center}

\end{center}

The role of adhesion in contact mechanics has seen an explosion of interest in
recent years, due to the enormous interest in soft materials technology,
nano-systems, cell adhesion, and the understanding of bio-attachments and the
idea to imitate their solutions (Creton \textit{et al.}, 1996, Kendall, 2001,
Kendall \textit{et al.}, 2010, Autumn \textit{et al.}, 2002). Ciavarella
\textit{et al.}(2018) discuss some aspects of various methods of solution,
stemming from the seminal paper of Johnson, Kendall, and Roberts (JKR, 1971)
who introduced an energy balance calculation like that of Griffith in fracture
mechanics (Maugis, 2013). The presence of surface roughness is so important
that for a long time it made impossible to measure adhesion between hard
materials, until JKR experimented on rubber, and Fuller and Tabor (1975) were
able to first measure the role of roughness. For nominally flat bulk solids,
it appears that the main solution to maintain stickiness is to reduce the
elastic modulus, as already suggested by the empirical Dahlquist (1969a,
1969b) criterion, which sets the threshold at the elastic Young modulus of
about $1$ $\mathrm{MPa}$.

Fuller and Tabor (1975) theory is based on asperities and its adhesion
parameter contains the {mean} asperity radius, which is not well defined for
"fractal" surfaces as today we consider, for which the "stickiness" would tend
to zero if we included extremely small wavelengths. Various other theories
have been proposed more recently (see the review by Ciavarella \textit{et
al.}(2018) for a general presentation), and there is debate still about the
applicability of each. Numerical solutions have clarified some aspects, and a
remarkable effort was made with the state-of-the-art M\"{u}ser's recent
`Contact Challenge' \ (M\"{u}ser \textit{et al.}, 2018, Ciavarella, 2018b).
However, they typically describe surfaces with PSDs spanning only about three
decades --- e.g. nanometer to micrometer scales, similarly to Pastewka and
Robbins (2014). Instead, the real "broadness" of the band of roughness is
likely to span many more decades of wavelength.

Indeed, in a very interesting recent paper by Dalvi \textit{et al.} (2019),
the authors describe topography across more than seven orders of magnitude,
including down to the \AA ngstr\"{o}m-scale, and the Power Spectrum Density
(PSD) follows almost a power law despite the broadness of the band (see fig.S2
which gives the 2D isotropic PSD). The "stickiness" criteria generated by
interpolating numerical results (Pastewka and Robbins, 2014, M\"{u}ser, 2016),
seem to depend critically on the truncation of the PSD, so further
investigation is important, and in particular, experiments. In these respects,
the very interesting recent paper by Dalvi \textit{et al.} (2019) reports
extensive adhesion measurements for soft elastic polydimethylsiloxane (PDMS)
hemispheres with elastic modulus ranging from 0.7 to 10 MPa in contact with
four different polycrystalline diamond substrates, and their careful
experimental effort corroborates ideas originally suggested by Persson (2002)
and Persson \&\ Tosatti (2002) inspired by the JKR energy balance concepts
(Johnson, Kendall and Roberts, 1971) of fracture mechanics applied to adhesion
of elastic bodies. We shall therefore further elaborate on the Persson and
Tosatti's ideas, and compare the results with other recent criteria, in
particular those proposed by Ciavarella (2018), and Violano \textit{et al.}
(2018). We shall find surprisingly universal results, despite the very
different origin of the various proposals we compare.

\subsection{Persson-Tosatti}

Persson (2002) and Persson \&\ Tosatti (2002) argue with a energy balance
between the state of full contact and that of complete loss of contact that
the effective energy available at pull-off with a rough interface is $\ $
\begin{equation}
\Delta\gamma_{eff}=\frac{A}{A_{0}}\Delta\gamma-\frac{U_{el}}{A_{0}}
\label{PerssonTosatti}%
\end{equation}
where $A$ is not the real contact area, but rather an area in full contact,
increased with respect to the nominal one $A_{0}$, because of an effect of
roughness-induced increase of contact area, $\frac{A}{A_{0}}>1$. Also,
$U_{el}$ is the elastic strain energy stored in the halfspace having roughness
with isotropic power spectrum $C\left(  q\right)  $ when this is squeezed
flat\footnote{Notice we use the original Persson's convention and notation for
$C\left(  q\right)  $ and not Dalvi et al. (2019) which is $C^{iso}\left(
q\right)  =4\pi^{2}C\left(  q\right)  $.}
\begin{equation}
\frac{U_{el}\left(  \zeta\right)  }{A_{0}}=\frac{\pi E^{\ast}}{2}\int_{q_{0}%
}^{q_{1}}q^{2}C\left(  q\right)  dq=E^{\ast}l\left(  \zeta\right)  \label{Uel}%
\end{equation}
where we have integrated over wavevectors in the range $q_{0}$, $q_{1}$, and
$E^{\ast}=E/\left(  1-\nu^{2}\right)  $ is the plane strain elastic modulus,
where $\nu$ is Poisson's ratio. We have introduced in (\ref{Uel}) a length
scale $l\left(  \zeta\right)  $ where $\zeta=q_{1}/q_{0}$ is the so called
"magnification". The elastic energy $U_{el}\left(  \zeta\right)  $ is
unbounded for surfaces with fractal dimension $D\geq2.5$, in the fractal limit
$\zeta\rightarrow\infty$ (see Ciavarella \textit{et al.}, 2018) so this theory
would predict that such surfaces could never adhere, even for arbitrarily
small rms height $h_{rms}$. This result may be in contrast with the theory by
Joe \textit{et al.} (2017, 2018), and should be further investigated. By
contrast, for $D<2.5$ the energy converges in the fractal limit $\zeta
\rightarrow\infty$ and hence full contact is expected to be possible
regardless of $h_{rms}$. Also, the simple theory has been shown to be a
reasonable approximation experimentally by Dalvi \textit{et al.} (2019). We
shall return on the Persson-Tosatti's idea to derive a stickiness criterion later.

\subsection{BAM theory}

The BAM model (Ciavarella, 2017) takes its inspiration from the DMT\ solution
for a single sphere (Derjaguin \textit{et al., }1975), completely different
from the JKR energy approach, but makes a geometric interpretation of it.
Hence, it doesn't follow any of the classical DMT calculations (neither the
thermodynamic method, nor the sum of adhesive forces in separation regions of
the adhesionless solution as done by Persson and Scaraggi, 2014). BAM assumes
the simplified Maugis-Dugdale force-separation law with a given interface
energy $\Delta\gamma$, for which the tensile stress is defined as a function
of gap $u$ as%

\begin{equation}%
\begin{tabular}
[c]{ll}%
$\sigma_{ad}\left(  u\right)  =\sigma_{0},$ & $u\leq\epsilon$\\
$\sigma_{ad}\left(  u\right)  =0,$ & $u>\epsilon$%
\end{tabular}
\ \ \ \qquad\qquad\qquad\qquad\label{maugis}%
\end{equation}
where $\sigma_{0}=\frac{\sigma_{th}}{16/\left(  9\sqrt{3}\right)  }%
\simeq\sigma_{th}$ (the theoretical strength of the material, for a
crystalline solid, and anyway the peak of tensile stress in a true
Lennard-Jones potential), $\epsilon$ is the range of attraction, and
$\Delta\gamma=\sigma_{0}\epsilon$. \ BAM makes an independent estimate for the
repulsive and adhesive components of the load. It has the big advantage to be
very simple to implement, particularly for rough surfaces, as it results in
closed form equations. The attractive area $A_{\mathrm{{ad}}}$ is defined as
\begin{equation}
A_{\mathrm{{att}}}(\Delta)\approx B(\Delta+\epsilon)-B(\Delta)\;,
\label{BAM01}%
\end{equation}
where $B(\Delta)$ is the classical \textit{bearing area,} namely the area over
which the bodies taken as rigid, would interpenetrate each other when moved
together through a distance $\Delta$. For a Gaussian nominally flat surface,
this results in
\begin{equation}
\frac{A_{ad}}{A_{0}}=\frac{1}{2}\left[  Erfc\left(  \frac{\overline
{u}-\epsilon}{\sqrt{2}h_{rms}}\right)  -Erfc\left(  \frac{\overline{u}}%
{\sqrt{2}h_{rms}}\right)  \right]
\end{equation}
where $\overline{u}$ is the mean separation of the surfaces, $h_{rms}$ is rms
amplitude of roughness. The total force is obtained by superposition of the
repulsive pressure at indentation $\Delta$ which is easily obtained with
Persson's theory (Persson, 2007) which, for the simplest power law PSD, and
$D\simeq2.2$ gives%
\begin{equation}
\frac{p_{\mathrm{{rep}}}\left(  \overline{u}\right)  }%
{E^{\raisebox{0.7mm}{$ *$}}}\simeq q_{0}h_{\mathrm{{rms}}}\exp\left(
\frac{-\overline{u}}{\gamma h_{\mathrm{{rms}}}}\right)  \label{Persson1}%
\end{equation}
where $\gamma\simeq0.5$ is a corrective factor. Therefore, summing up
repulsive\ (\ref{Persson1}) and attractive ($\sigma_{0}A_{\mathrm{{ad}}%
}(\overline{u})$) contributions, BAM gives
\begin{equation}
\frac{\sigma\left(  \overline{u}\right)  }{\sigma_{0}}\simeq q_{0}h_{rms}%
\frac{E^{\ast}}{\sigma_{0}}\exp\left(  \frac{-\overline{u}}{\gamma h_{rms}%
}\right)  -\frac{1}{2}\times\left[  Erfc\left(  \frac{\overline{u}-\epsilon
}{\sqrt{2}h_{rms}}\right)  -Erfc\left(  \frac{\overline{u}}{\sqrt{2}h_{rms}%
}\right)  \right]  \label{magic}%
\end{equation}
which obviously results in a pull off finding the minimum as a function of
$\overline{u}$. Notice that $\frac{E^{\ast}}{\sigma_{0}}\simeq\frac{E^{\ast}%
}{\Delta\gamma/\epsilon}=\frac{\epsilon}{l_{a}}$ where $l_{a}=\Delta
\gamma/E^{\ast}$ defines a characteristic adhesion length which for the
typical Lennard Jones description of an interface between crystals of the same
material is $l_{a}\simeq0.05\epsilon$. The theoretical strength in this case,
$\sigma_{0}=\frac{\Delta\gamma}{\epsilon}=\frac{l_{a}E^{\ast}}{\epsilon
}=0.05E^{\ast}$. However, when considering contamination, one can estimate
that $l_{a}$ is reduced by orders of magnitude. The results show that the
pull-off traction is principally determined by $h_{rms},q_{0}$ and upon
increasing the "magnification" of the surface, $\zeta=q_{1}/q_{0}$, converges
rapidly, as in the adhesionless load-separation relation (\ref{Persson1}). We
shall return later on BAM\ to derive a stickiness criterion also from it.

\subsection{Violano \textit{et al.} criterion}

{Inspired by some concepts originally introduced by Pastewka and Robbins
(2014), namely about the presence of a boundary layer near the edge of contact
where gaps could be described by universal asymptotic expressions, Violano
\textit{et al.} (2018) obtained the probability density function of gaps with
Persson and Scaraggi's DMT theory (Persson and Scaraggi, 2014, see also
Afferrante \textit{et al.}(2018)) and found that it converges with increasing
magnification $\zeta$, thus, in the fractal limit, any DMT theory should not
depend on the PSD wavenumber cutoff $q_{1}$ --- thereby showing a different
extrapolation to broad band roughness than Pastewka and Robbins (2014) who had
numerically explored only up to }$\zeta\simeq10^{3}$.{ Violano and co-authors
showed that the area-load slope, at the origin (which becomes vertical when we
move from sticky to unsticky), depends in a pure power law PSD only on
well-defined macroscopic quantities, such as $h_{rms}$ and the lowest
wavenumber $q_{0}$, and in particular that for low fractal dimension
$(D\simeq2.2)$ rough surfaces stick for }%

\begin{equation}
\frac{h_{rms}}{\epsilon}<\left(  \frac{9}{4}\frac{\sigma_{0}/E^{\ast}%
}{\epsilon q_{0}}\right)  ^{3/5} \label{VIOLANO3}%
\end{equation}
which we are going to use for comparative purposes.

\section{\bigskip New stickiness criteria}

\subsection{A new Persson-Tosatti stickiness criterion}

Let us start from obtaining a stickiness criterion from the Persson-Tosatti's
idea of the effective surface energy (\ref{PerssonTosatti}). If we take a
typical power law PSD $C\left(  q\right)  =Zq^{-2\left(  1+H\right)  }$ for
$q>q_{0}=\frac{2\pi}{\lambda_{L}}$, where $H$ is the Hurst exponent (equal to
$3-D$ where $D$ is the fractal dimension of the surface), the integral of the
full contact energy (\ref{Uel}) depends on whether $H>0.5$ or not.
Specifically, as $Z=\frac{H}{2\pi}\left(  \frac{h_{0}}{q_{0}}\right)
^{2}\left(  \frac{1}{q_{0}}\right)  ^{-2\left(  H+1\right)  }$ where
$h_{0}^{2}=2h_{rms}^{2}$ (see again Persson, 2002), for $H\neq0.5$
\begin{equation}
l\left(  \zeta\right)  =\frac{\pi}{2}\int_{q_{0}}^{q_{1}}q^{2}C\left(
q\right)  dq=\frac{\pi Z}{2}\int_{q_{0}}^{q_{1}}q^{-2H}dq=\pi\frac{h_{rms}%
^{2}}{\lambda_{L}}H\frac{\zeta^{-2H+1}-1}{-2H+1}%
\end{equation}

For the usual case of $H>0.5$ (low $D$) (see Persson 2014) the integral
converges quickly, is relatively insensitive to high wavevector truncation and
indeed for practical purposes we can use the limit value%
\begin{equation}
l\left(  \infty\right)  _{lowD}=\pi\frac{h_{rms}^{2}}{\lambda_{L}}\frac
{H}{2H-1}%
\end{equation}
which shows the energy is mainly stored in the long wavelength components.
Summarizing, and neglecting the effect of the term $A/A_{0}$, we can simplify
the effective surface energy (\ref{PerssonTosatti}) as
\begin{equation}
\Delta\gamma_{eff}=\Delta\gamma-E^{\ast}\pi\frac{h_{rms}^{2}}{\lambda_{L}%
}\frac{H}{2H-1} \label{PT1}%
\end{equation}
We can then obtain a new "Persson-Tosatti" stickiness criterion, by imposing
$\Delta\gamma_{eff}=0$ in (\ref{PT1}) obtaining in terms of roughness
amplitude, the condition%
\begin{equation}
h_{rms}<\sqrt{\frac{\Delta\gamma}{E^{\ast}}\lambda_{L}\frac{2H-1}{\pi H}}
\label{PTcriterion}%
\end{equation}
which we shall compare with other criteria.

\subsection{A new BAM stickiness criterion}

We have not obtained in the original BAM\ paper (Ciavarella, 2018), a true
criterion for stickiness, and this does not seem to be obtained in closed
form. One can obtain from eqt (\ref{magic}) the decay of the pull-off tension
(fig.1) as a function of rms roughness amplitude. Given the abrupt decay in
pull-off values, stickiness is defined (for example) when -$\sigma_{\min
}/\sigma_{0}=10^{-8}$ finding this by numerical routines. Moreover, defining
the threshold from the exact minimum of the tension-mean gap curve (solid
lines in Fig.1), or defining it from the curves obtained at $\overline
{u}/\epsilon=2$ (dashed lines) is the same as clearly demonstrated by the
Fig.1, so that one can find the threshold for stickiness also directly from
solving the following equation $f\left(  \lambda_{L}/\epsilon,\left(
h_{rms}/\epsilon\right)  _{thresh},l_{a}/\epsilon\right)  =10^{-8}$ where
\begin{align}
f\left(  \lambda_{L}/\epsilon,\left(  h_{rms}/\epsilon\right)  _{thresh}%
,l_{a}/\epsilon\right)   &  =\frac{2\pi}{l_{a}/\epsilon}\frac{\left(
h_{rms}/\epsilon\right)  _{thresh}}{\lambda_{L}/\epsilon}\exp\left(  \frac
{-2}{\gamma\left(  h_{rms}/\epsilon\right)  _{thresh}}\right) \nonumber\\
&  -\frac{1}{2}\times\left[  Erfc\left(  \frac{1}{\sqrt{2}\left(
h_{rms}/\epsilon\right)  _{thresh}}\right)  -Erfc\left(  \frac{2}{\sqrt
{2}\left(  h_{rms}/\epsilon\right)  _{thresh}}\right)  \right]  \label{thresh}%
\end{align}

\begin{center}
$%
\begin{array}
[c]{c}%
%TCIMACRO{\FRAME{itbpF}{5.0548in}{3.1194in}{0in}{}{}{j1.eps}%
%width 5.0548in;  height 3.1194in;  depth 0in;  original-width 5.0004in;
%original-height 3.0753in;  cropleft "0";  croptop "1";  cropright "1";
%cropbottom "0";  filename 'j1.eps';file-properties "XNPEU";}} }%
%BeginExpansion
{\includegraphics[
height=3.1194in,
width=5.0548in
]%
{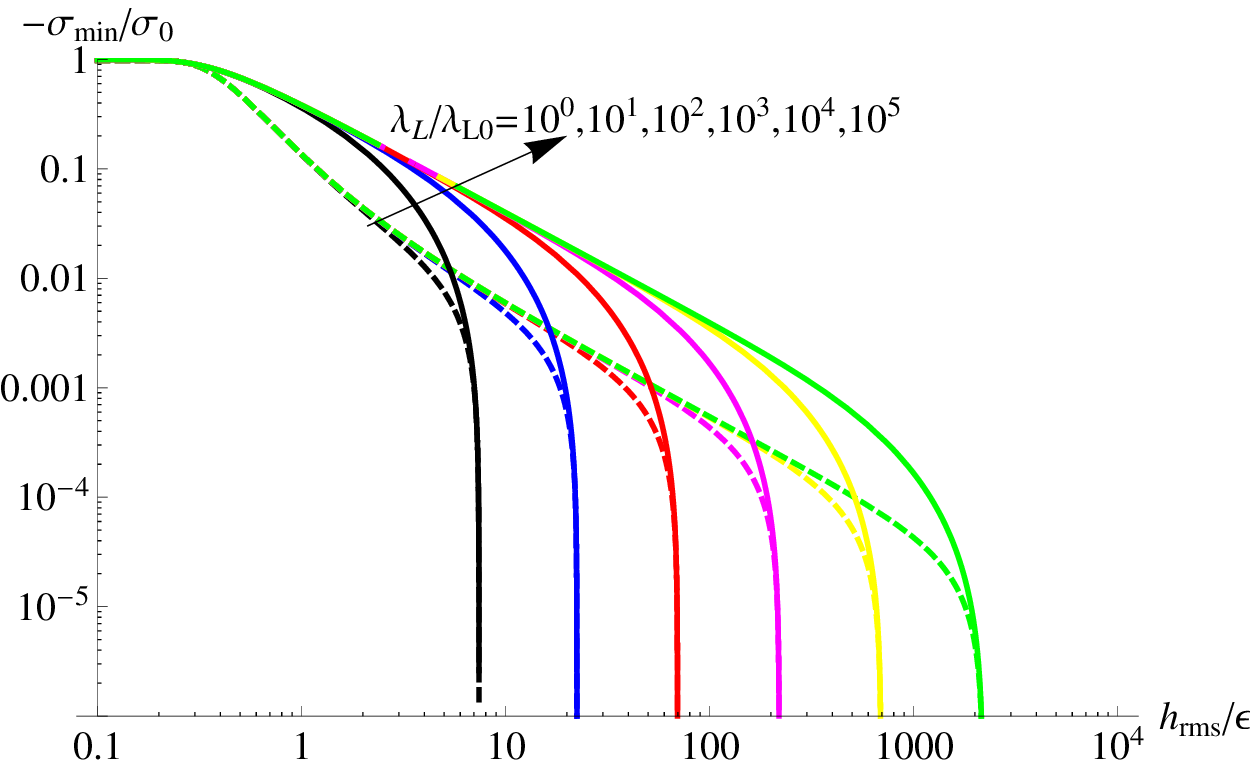}%
}
%EndExpansion
\\
\end{array}
$

Fig.1- Curves of decay of pull-off normalized pressure -$\sigma_{\min}%
/\sigma_{0}$ as a function of normalized rms roughness amplitude
$h_{rms}/\epsilon$ (solid lines represent the true pull-off point, while
dashed lines represent an approximation computed at $\overline{u}/\epsilon
=2$). Here, the reference long wavelength cutoff $\lambda_{L0}=\frac{q_{0}%
}{2\pi}=2048\epsilon$ and the curves shift to the right with increasing
$\lambda_{L}/\lambda_{L0}=10^{0},10^{1}...10^{5}$.
\end{center}

Hence, we can explore the threshold $\left(  h_{rms}/\epsilon\right)
_{thresh}$ so obtained in Fig.2, where solid lines represent the actual
solutions to eqt. (\ref{thresh}), and dashed lines represent power law
approximations of the type%
\begin{equation}
\left(  h_{rms}/\epsilon\right)  _{thresh}=\alpha\left(  \frac{l_{a}}%
{\epsilon}\right)  \left(  \lambda_{L}/\lambda_{L0}\right)  ^{1/2}
\label{hpower}%
\end{equation}
which provides a very reasonable fit over various orders of magnitude of
$\left(  \lambda_{L}/\lambda_{L0}\right)  $, taking as reference $\lambda
_{L0}/\epsilon=2048$. Supposing $\epsilon$ in the Angstrom range, 10$^{-10}m$,
$\lambda_{L}/\lambda_{L0}=10^{7}$ means we are effectively plotting up to $mm$
range, similarly to the broadness of roughness measured in Davli \textit{et
al.} (2019).

\begin{center}
$%
\begin{array}
[c]{c}%
%TCIMACRO{\FRAME{itbpF}{5.0548in}{3.1038in}{0in}{}{}{j2.eps}%
%width 5.0548in;  height 3.1038in;  depth 0in;  original-width 5.0004in;
%original-height 3.0597in;  cropleft "0";  croptop "1";  cropright "1";
%cropbottom "0";  filename 'j2.eps';file-properties "XNPEU";}} }%
%BeginExpansion
{\includegraphics[
height=3.1038in,
width=5.0548in
]%
{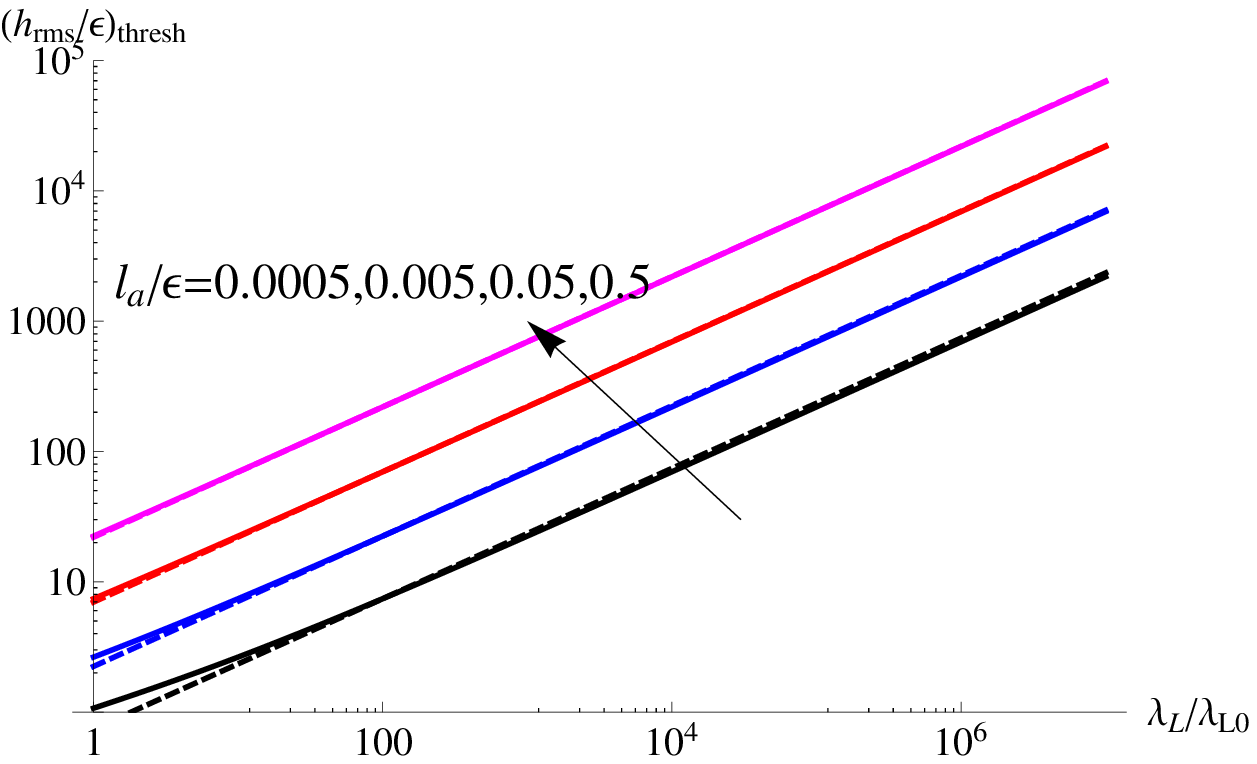}%
}
%EndExpansion
\\
\end{array}
$

Fig.2- The threshold for stickiness $\left(  h_{rms}/\epsilon\right)
_{thresh}$ as a function of $\lambda_{L}/\lambda_{L0}$ across 7 orders of
magnitude where $\lambda_{L0}=\frac{q_{0}}{2\pi}=2048\epsilon$ shows extremely
good power law behaviour for all values of $\left(  \frac{l_{a}}{\epsilon
}\right)  $. Solid lines represent the actual solutions to eqt. (\ref{thresh}%
), and dashed lines represent power law approximations (\ref{hpower})
\end{center}

The constant $\alpha\left(  \frac{l_{a}}{\epsilon}\right)  $ is further
studied in Fig.3, showing that even this quantity has a very good power law
fit across many orders of magnitude of $\frac{l_{a}}{\epsilon}$, namely we can
write (shown as dashed line)%
\begin{equation}
\alpha\left(  \frac{l_{a}}{\epsilon}\right)  \simeq35\left(  \frac{l_{a}%
}{\epsilon}\right)  ^{0.5} \label{alfa}%
\end{equation}

\begin{center}
$%
\begin{array}
[c]{c}%
%TCIMACRO{\FRAME{itbpF}{5.056in}{3.2179in}{0in}{}{}{j3.eps}%
%width 5.056in;  height 3.2179in;  depth 0in;  original-width 4.8003in;
%original-height 3.0461in;  cropleft "0";  croptop "1";  cropright "1";
%cropbottom "0";  filename 'j3.eps';file-properties "XNPEU";}} }%
%BeginExpansion
{\includegraphics[
height=3.2179in,
width=5.056in
]%
{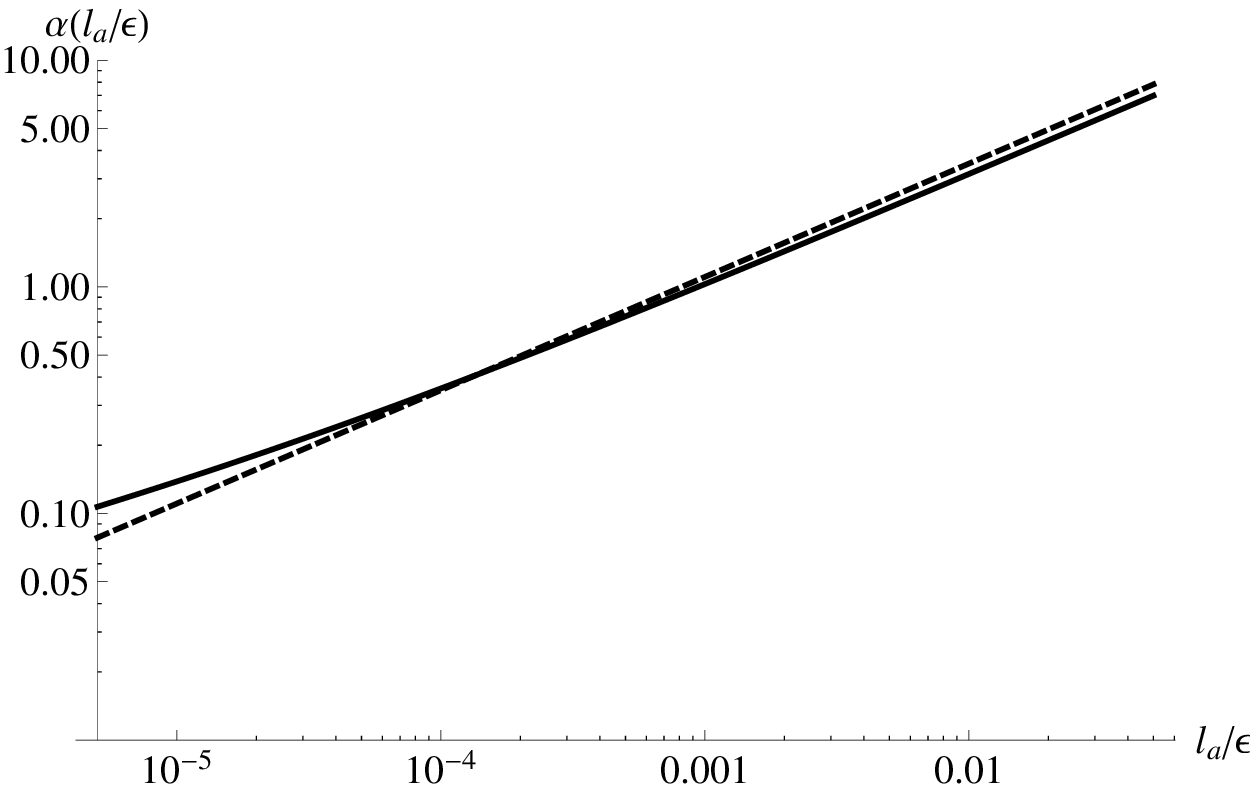}%
}
%EndExpansion
\\
\end{array}
$

Fig.3- The constant of proportionality $\alpha\left(  \frac{l_{a}}{\epsilon
}\right)  $ as a function of $\frac{l_{a}}{\epsilon}$ across 4 orders of
magnitude shows also extremely good power law behaviour (\ref{alfa}).
\end{center}

Summarizing, using (\ref{hpower}, \ref{alfa}), the dependence on $\epsilon$
disappears and we have obtained \bigskip for stickiness%
\begin{equation}
h_{rms}<\left(  0.6l_{a}\lambda_{L}\right)  ^{0.5} \label{bam-stickiness}%
\end{equation}
which we shall compare with the other criteria.

\section{\bigskip Comparison between the three stickiness criteria}

We have obtained, for the example case of a pure power law PSD roughness (for
the typical case of $H\simeq0.8$) that Persson-Tosatti and BAM predict
stickiness (\ref{PTcriterion}) (\ref{bam-stickiness}) with exactly the same
qualitative form
\begin{equation}
h_{rms}<\sqrt{\beta l_{a}\lambda_{L}}\ \label{PT-BAM}%
\end{equation}
where $\beta_{PT}=0.24$ and $\beta_{BAM}=0.6$ which are even
\textit{quantitatively} close --- even closer they will appear considering the
factor $A/A_{0}$ as we demonstrate in the discussion. The result is really
unexpected, given the two simple criteria are obtained with completely
different routes, one being a simple energy balance without considering
details of the contact mechanics, and the other a mix of Persson's solution
for repulsive pressure, and a geometrical estimate for the adhesive pressure.

Violano's DMT criterion instead (\ref{VIOLANO3}) contains a slightly different
qualitative dependence on material properties, since we can write it in the form%

\begin{equation}
h_{rms}<\epsilon^{-1/5}\left(  0.36l_{a}\lambda_{L}\right)  ^{3/5}%
\end{equation}
\bigskip which shows the product \ $l_{a}\lambda_{L}$, instead of the power
1/2, is raised to the power 3/5, and this is due to a weak apparent dependence
on the range of attractive forces $\epsilon$.\ 

We can rewrite all three criteria by introducing the parameter $\epsilon$
(which is purely a normalization factor for Persson-Tosatti and BAM, whereas
it is a true dependent parameter for Violano's), in the form{ }%
\begin{align}
\frac{h_{rms}}{\epsilon}  &  <\sqrt{0.24\frac{l_{a}}{\epsilon}\frac
{\lambda_{L}}{\epsilon}}\text{\qquad;\qquad\ Persson-Tosatti}\label{c1}\\
\frac{h_{rms}}{\epsilon}  &  <\sqrt{0.6\frac{l_{a}}{\epsilon}\frac{\lambda
_{L}}{\epsilon}}\text{\qquad;\qquad\ BAM}\label{c2}\\
\frac{h_{rms}}{\epsilon}  &  <\left(  0.358\frac{l_{a}}{\epsilon}\frac
{\lambda_{L}}{\epsilon}\right)  ^{3/5}\text{\qquad;\qquad\ Violano} \label{c3}%
\end{align}
and a comparison is shown in Fig.4, where Persson-Tosatti is reported in black
solid line, BAM as blue solid line, and\ Violano as red solid line. Clearly,
considering the three criteria have so different origin, it is remarkable that
they give so close results.

\begin{center}
$%
\begin{array}
[c]{cc}%
%TCIMACRO{\FRAME{itbpF}{5.056in}{3.1324in}{0in}{}{}{j4.eps}%
%width 5.056in;  height 3.1324in;  depth 0in;  original-width 4.8003in;
%original-height 2.9639in;  cropleft "0";  croptop "1";  cropright "1";
%cropbottom "0";  filename 'j4.eps';file-properties "XNPEU";}} }%
%BeginExpansion
{\includegraphics[
height=3.1324in,
width=5.056in
]%
{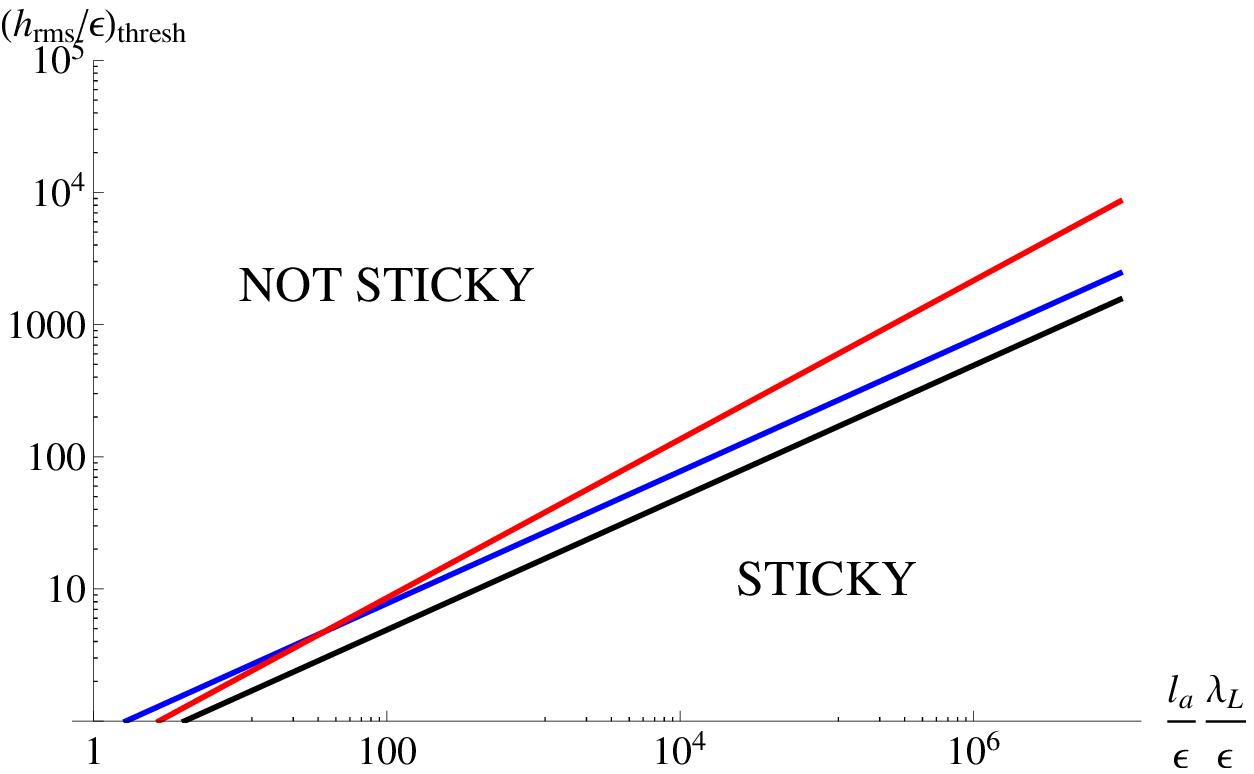}%
}
%EndExpansion
&
\end{array}
$

Fig.4. A comparison of the three derived stickiness criteria: Persson-Tosatti
(black line) (\ref{c1}), BAM (blue solid line) (\ref{c2}) and Violano (red
line) (\ref{c3}), in terms of the rms amplitude of roughness. All results are
for $H=0.8$ ($D=2.2$)
\end{center}

\bigskip

Further comparisons between the different criteria would require very
sophisticated experiments, and those of Davli \textit{et al.} (2019) are
relevant to a spherical geometry, whereas all three criteria are in principle
obtained from theories on contact about nominally flat surfaces. Writing
{$h_{rms}\simeq\sqrt{\pi Z/H}q_{0}^{-H}$, taking the new Persson-Tosatti
criterion in the reelaborated following form }%
\begin{equation}
Z<1.16\frac{\Delta\gamma}{E^{\ast}}\left(  \frac{1}{\lambda_{L}}\right)
^{0.6}%
\end{equation}
we find for $\Delta\gamma=37mJ/m^{2}$ and $E^{\ast}=0.7-10MPa$, with
$\lambda_{L}\simeq10mm$, while from the plots in Davli \textit{et al.} (2019),
we can estimate an approximate power law with $Z\simeq C\left(  q\right)
q^{3.6}\simeq\allowbreak2.5\times10^{-10}m^{0.4}<1.16\frac{\Delta\gamma
}{E^{\ast}}\left(  \frac{1}{\lambda_{L}}\right)  ^{0.6}=\allowbreak
6.8\times10^{-8}m^{0.4}$, all data should be sticky, as it appears the case.

\section{Discussion}

\subsection{Persson-Tosatti and BAM even closer when considering surface area
increase}

Dalvi \textit{et al.} (2019) make a discussion about the term true surface
area, which they estimate as
\[
\frac{A\left(  \zeta\right)  }{A_{0}}=1+\frac{\sqrt{\pi}}{2}h_{rms}^{\prime
2}\exp\left(  \frac{1}{h_{rms}^{\prime2}}\right)  Erfc\left(  \frac{1}%
{h_{rms}^{\prime}}\right)  /h_{rms}^{\prime}%
\]
which complicates the model slightly. This terms will modify the
Persson-Tosatti criterion introducing a magnification-dependence
\begin{equation}
h_{rms}<\sqrt{\frac{A\left(  \zeta\right)  }{A_{0}}\frac{\Delta\gamma}%
{E^{\ast}}\lambda_{L}\frac{2H-1}{\pi H}}%
\end{equation}
For example, let us consider the usual case $H=0.8$, this will make the
stickiness criterion look like
\begin{equation}
h_{rms}<\sqrt{\beta_{PT}\left(  \zeta\right)  l_{a}\lambda_{L}}\
\end{equation}
where the $\beta_{PT}=0.24\frac{A\left(  \zeta\right)  }{A_{0}}$ \ term
increase only with a quite large $h_{rms}^{\prime}\ $\ as plotted in Fig.5.
Incidentally this will get even closer to the $\beta_{BAM}=0.6$ despite for a
different reason. An almost perfect coincidence between Persson-Tosatti
criterion and BAM will occur at some $h_{rms}^{\prime}\simeq2.51$ which is a
realistic value for a limit true slope --- above which many other assumptions
of linear elasticity, of geometrical description of atomic structures etc.
would be violated.

\begin{center}
$%
\begin{array}
[c]{cc}%
%TCIMACRO{\FRAME{itbpF}{5.056in}{3.1905in}{0in}{}{}{j5.eps}%
%width 5.056in;  height 3.1905in;  depth 0in;  original-width 4.8003in;
%original-height 3.0195in;  cropleft "0";  croptop "1";  cropright "1";
%cropbottom "0";  filename 'j5.eps';file-properties "XNPEU";}} }%
%BeginExpansion
{\includegraphics[
height=3.1905in,
width=5.056in
]%
{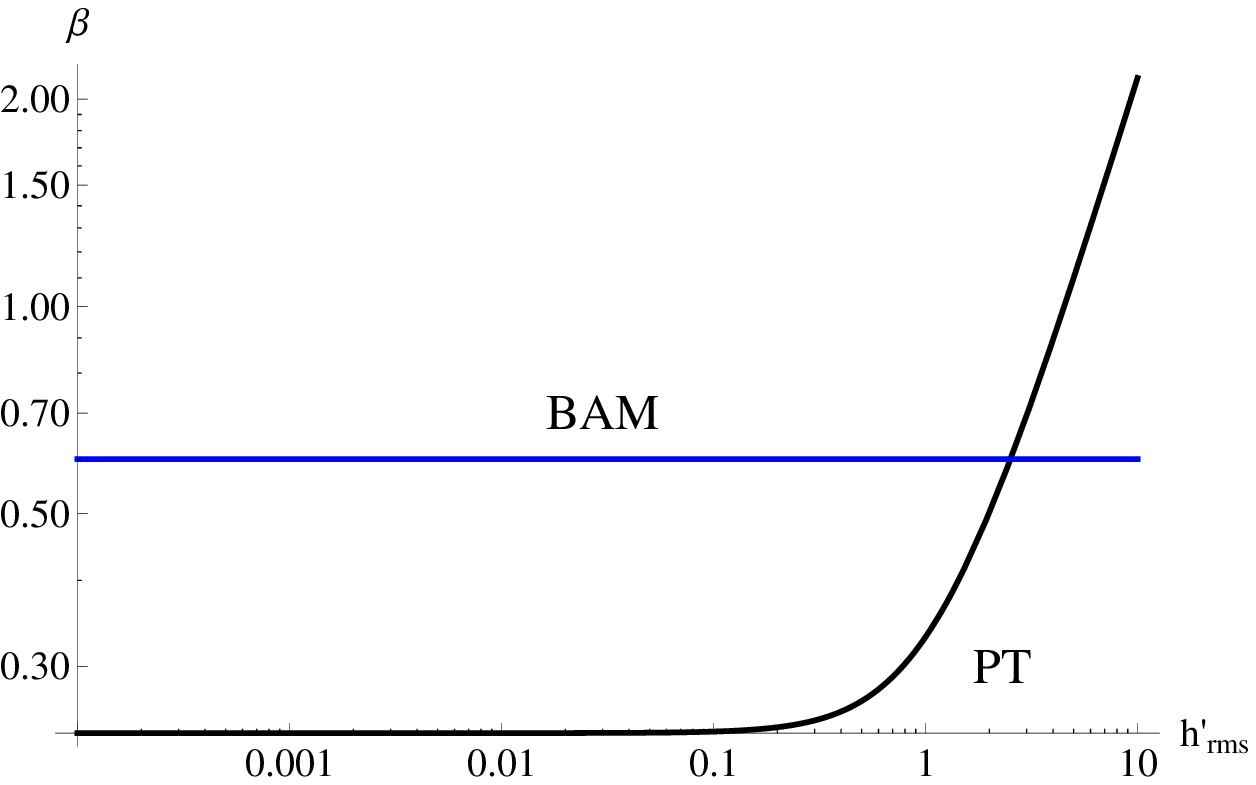}%
}
%EndExpansion
&
\end{array}
$

Fig.5. The coefficient $\beta$ in the Persson-Tosatti and BAM stickiness
criteria $h_{rms}<\sqrt{\beta\left(  \zeta\right)  l_{a}\lambda_{L}}$ as a
function of the rms slope
\end{center}

The correction $\frac{A\left(  \zeta\right)  }{A_{0}}$ was discussed also in
Persson's own later experiments (Peressadko \textit{et al.}, 2005).

\subsection{Other criteria}

We haven't so far commented on the Pastewka and Robbins (2014) stickiness
criterion, which we can rewrite it in the form (for power law tail of the
PSD)
\begin{equation}
\frac{\epsilon}{l_{a}}\left(  \frac{h_{rms}}{\epsilon}\right)  ^{2/3}<\frac
{3}{2}\frac{a_{V}\left(  \zeta\right)  }{q_{0}h_{rms}} \label{VIOLANO1}%
\end{equation}
where
\begin{equation}
\left[  a_{V}\left(  \zeta\right)  \right]  _{PR}=1.4622q_{0}h_{rms}\left(
\frac{h_{rms}^{\prime\prime2}}{h_{rms}^{\prime7}}\right)  ^{1/3}h_{rms}^{2/3}%
\end{equation}

Now for power law PSD, estimating $h_{rms}=\sqrt{{m_{0}}}${$=\sqrt{2\pi
Zq_{0}^{-2H}\left(  \frac{\zeta^{-2H}-1}{-2H}\right)  }\simeq\sqrt{\frac{\pi
Z}{H}}q_{0}^{-H}$ and} $h_{rms}^{\prime}=\sqrt{2m_{2}}\simeq\sqrt{\frac{\pi
Z}{1-H}}q_{1}^{1-H},h_{rms}^{\prime\prime}=\sqrt{8m_{4}/3}\simeq\sqrt
{2\frac{\pi Z}{4-2H}}q_{1}^{2-H}$ , where $m_{0},m_{2},m_{4}\ $\ are spectral
moments of the PSD, we obtain
\begin{equation}
\left[  a_{V}\left(  \zeta\right)  \right]  _{PR}=1.4622\frac{\left(
1-H\right)  ^{7/6}}{\left(  2-H\right)  ^{1/3}H^{5/6}}\zeta^{\frac{5}{3}H-1}%
\end{equation}
which for $H=0.8$ leads to $\left[  a_{V}\left(  \zeta\right)  \right]
_{PR}=0.253$ and therefore using again (\ref{VIOLANO1})%
\begin{equation}
\frac{h_{rms}}{\epsilon}<\left(  0.06\frac{l_{a}}{\epsilon}\frac{\lambda_{L}%
}{\epsilon}\right)  ^{3/5}\zeta^{1/5}%
\end{equation}

Now, this shows a (weak) dependence on magnification, which remains even for
high $\zeta$, unlike the other criteria. If we plot the Pastewka and Robbins
criterion as in Fig.4, we obtain fig.6 for $\zeta=10^{3},10^{4},...10^{7}$
(where we add to the three previous lines dashed lines corresponding to
Pastewka and Robbins criterion for increasing $\zeta$, increasing as indicated
by arrow). It is evident that the PR criterion corresponds very closely to the
Violano criterion for low $\zeta<1000$ (which is where it was obtained), but
departs for higher $\zeta$. Hence, in practical cases shown by Davli
\textit{et al.} (2019) who have $\zeta\simeq10^{7}$, it is safer to use the
other three criteria which all do not show this dependence, probably found
spuriously from the limited numerical experiments. More specifically, we don't
really need to measure surface roughness down to atomic scale, since the three
criteria (Persson-Tosatti, BAM and Violano), all do not require very precise
informations about small scale details to be defined.

\begin{center}
$%
\begin{array}
[c]{cc}%
%TCIMACRO{\FRAME{itbpF}{5.056in}{3.1324in}{0in}{}{}{j6.eps}%
%width 5.056in;  height 3.1324in;  depth 0in;  original-width 4.8003in;
%original-height 2.9639in;  cropleft "0";  croptop "1";  cropright "1";
%cropbottom "0";  filename 'j6.eps';file-properties "XNPEU";}} }%
%BeginExpansion
{\includegraphics[
height=3.1324in,
width=5.056in
]%
{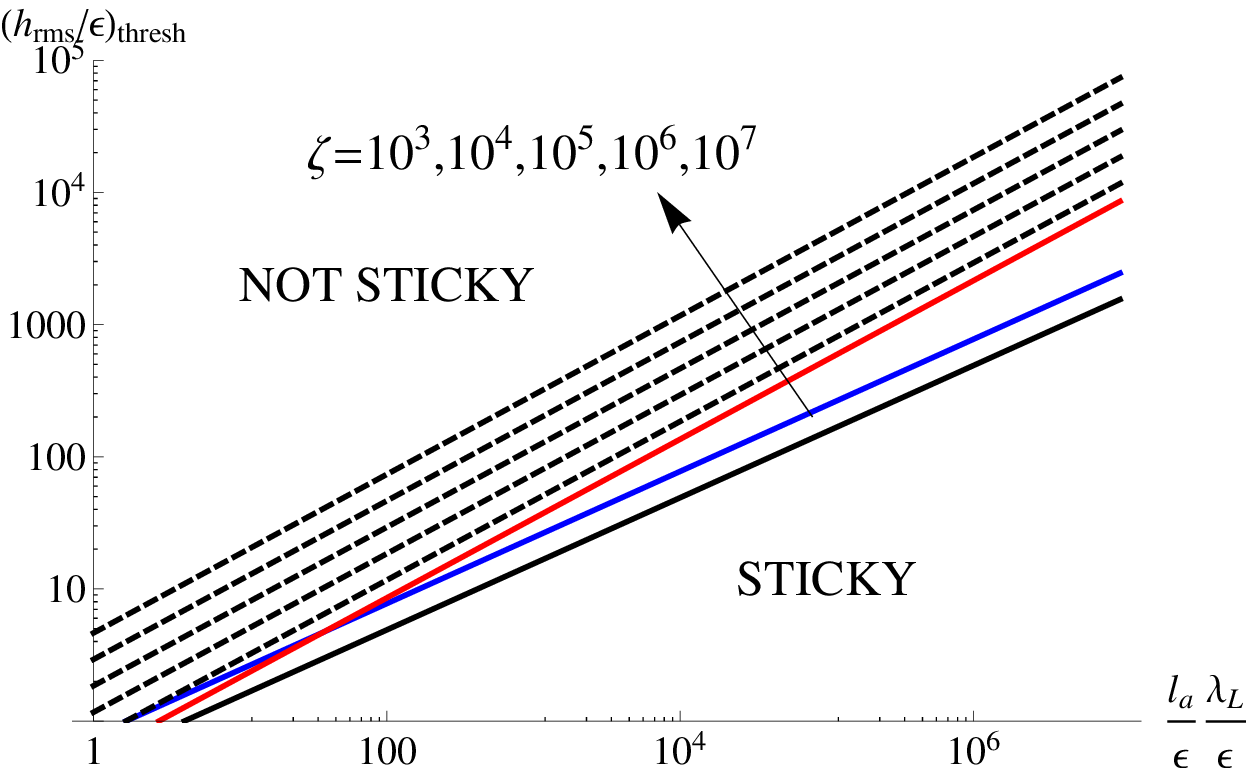}%
}
%EndExpansion
&
\end{array}
$

Fig.6. A further comparison of the three derived stickiness criteria,
Persson-Tosatti (black line), BAM (blue solid line) and Violano (red line),
with a fourth criterion, that of Pastewka and Robbins (2014) which was
obtained only by interpolation of numerical results for $\zeta<1000$ --- this
shows that there is good agreement even with the PR criterion but only for low
magnifications, and beyond this (as in practical cases shown by Davli
\textit{et al.} who have $\zeta\simeq10^{7}$), it is safer to use the other
three criteria.
\end{center}

\bigskip

Similarly to Pastewka and Robbins (2014), M\"{u}ser (2016) also defines a
stickiness criterion interpolating numerical results, defines a "dimensionless
surface energy",
\begin{equation}
\Delta\gamma_{rss}=\frac{\Delta\gamma}{E^{\ast}}\frac{\tanh\left(  \mu
_{T}\right)  }{\left(  h_{rms}^{\prime}\right)  ^{3}}%
\end{equation}
where $h_{rms}^{\prime}$ is the root mean-square gradient of the surface,
$\tanh$ is introduced as an empirical fitting between the "correct"
asymptotics in the two limits of small and large Tabor generalized
coefficients $\mu_{T}$ (see M\"{u}ser (2016) for details). For the power law
PSD spectrum $h_{rms}^{\prime}\sim\zeta^{1-H}$, $h_{rms}^{\prime\prime}%
\sim\zeta^{2-H},$ and hence $\tanh\left(  \mu_{T}\right)  \sim$ $\left(
h_{rms}^{\prime\prime}\right)  ^{-1/3}$ so that
\begin{equation}
\Delta\gamma_{rrs}\sim\zeta^{\left(  2-H\right)  \frac{2}{3}-3+3H}\sim
\zeta^{\left(  7H-5\right)  /3},\qquad\zeta\rightarrow\infty
\end{equation}
which means that for $H=0.8$ that $\Delta\gamma_{rrs}\rightarrow\zeta^{0.2}$,
and again this shows a magnification dependence for all $\zeta$ similarly to
Pastewka and Robbins (2014), but in contrast with the Persson-Tosatti, BAM,
and Violano criteria.

\section{Conclusions}

We have obtained two new stickiness criteria, originated from the theories of
Persson-Tosatti, and from BAM. These two, which have completely different
origin (one being a simple energy balance concept, and the other a mix of
Persson's adhesiveless solution with a geometric estimate of adhesive forces),
together with the DMT criterion of Violano \textit{et al.} which in turn is
based on the elaborated DMT theory of Persson and Scaraggi, seem to differ
only by prefactors (Persson-Tosatti vs BAM), or by a small difference in the
power laws exponent, due to a weak apparent dependence on the range of
attractive forces, emerging in the Violano's criterion. However, all three
criteria show the main factors for stickiness are the low wavevector cutoff of
roughness, $q_{0}=\frac{2\pi}{\lambda_{L}}$, the rms amplitude of roughness
$h_{rms}$ and the ratio between the work of adhesion and the plane strain
Young modulus. We find this result rather surprising and hence a robust
indication now that small scale features (such as local slopes or curvature,
which are hard to define down to perhaps atomic scale) do not affect
stickiness. For adhesion to various levels of macroscopic roughness, the only
characteristic which can be changed easily is the elastic modulus, in
qualitative and quantitative agreement with Dahlquist criterion.

\section*{Acknowledgements}

MC acknowledges support from the Italian Ministry of Education, University and
Research (MIUR) under the program "Departments of Excellence" (L.232/2016).

\section{References}

Afferrante, L., Bottiglione, F., Putignano, C., Persson, B. N. J., \& Carbone,
G. (2018). Elastic Contact Mechanics of Randomly Rough Surfaces: An Assessment
of Advanced Asperity Models and Persson's Theory. Tribology Letters, 66(2), 75.

Autumn, K., Sitti, M., Liang, Y. A., Peattie, A. M., Hansen, W. R., Sponberg,
S., Kenny, T.W., Fearing, R., Israelachvili, J.N. \& Full, R. J. (2002).
Evidence for van der Waals adhesion in gecko setae. Proceedings of the
National Academy of Sciences, 99(19), 12252-12256

Ciavarella, M., Joe, J., Papangelo, A., Barber, JR. (2019) The role of
adhesion in contact mechanics. J. R. Soc. Interface, 16, 20180738

Ciavarella, M. (2018) A very simple estimate of adhesion of hard solids with
rough surfaces based on a bearing area model. Meccanica, 1-10. DOI 10.1007/s11012-017-0701-6

Ciavarella, M. (2018b). A Comment on \textquotedblleft Meeting the
Contact-Mechanics Challenge\textquotedblright\ by Muser et al.[1]. Tribology
Letters, 66(1), 37.

Creton C., Ciccotti M. (2016). Fracture and adhesion of soft materials: a
review. Reports on Progress in Physics, 79(4), 046601.

Dahlquist, C. A. in Treatise on Adhesion and Adhesives, R. L. Patrick (ed.),
Dekker, New York, 1969a ,2, 219.

Dahlquist, C., Tack, in Adhesion Fundamentals and Practice. 1969b, Gordon and
Breach: New York. p. 143-151.

Dalvi, S., Gujrati, A., Khanal, S. R., Pastewka, L., Dhinojwala, A., \&
Jacobs, T. D. (2019). Linking energy loss in soft adhesion to surface
roughness. arXiv preprint arXiv:1907.12491.

\bigskip Derjaguin, B. V., Muller V. M. \& Toporov Y. P. (1975). Effect of
contact deformations on the adhesion of particles. J. Colloid Interface Sci.,
53, pp. 314--325.

Fuller, K.N.G., Tabor, D., (1975), The effect of surface roughness on the
adhesion of elastic solids. Proc. R. Soc. Lond. A, 345(1642), 327-342.

Gao, H., Wang, X., Yao, H., Gorb, S., \& Arzt, E. (2005). Mechanics of
hierarchical adhesion structures of geckos. Mechanics of Materials, 37(2-3), 275-285.

Joe, J., Scaraggi, M., \& Barber, J. R. (2017). Effect of fine-scale roughness
on the tractions between contacting bodies. Tribology International, 111,
52--56. https://doi.org/10.1016/j.triboint.2017.03.001

Joe, J., Thouless, M.D. , Barber, J.R. (2018), Effect of roughness on the
adhesive tractions between contacting bodies, Journal of the Mechanics and
Physics of Solids, doi.org/10.1016/j.jmps.2018.06.005

\bigskip Johnson, K.L. , Kendall, K. , Roberts, A.D. (1971), Surface energy
and the contact of elastic solids. Proc R Soc Lond;A324:301--313. doi: 10.1098/rspa.1971.0141

Kendall K (2001) Molecular Adhesion and Its Applications: The Sticky Universe
(Kluwer Academic, New York)

Kendall K., Kendall M., Rehfeld F. (2010).Adhesion of Cells, Viruses and
Nanoparticles, Springer Dordrecht Heidelberg London New York.

Maugis, D. (2013). Contact, adhesion and rupture of elastic solids (Vol. 130).
Springer Science \& Business Media.

M\"{u}ser, MH., Dapp WB., Bugnicourt R., Sainsot P., Lesaffre, N., Lubrecht,
T.A., Persson BNJ \textit{et al.} (2017) "Meeting the contact-mechanics
challenge." Tribology Letters 65, no. 4, 118

M\"{u}ser, M. H. (2016). A dimensionless measure for adhesion and effects of
the range of adhesion in contacts of nominally flat surfaces. Tribology
International, 100, 41-47.

Pastewka, L., \& Robbins, M. O. (2014). Contact between rough surfaces and a
criterion for macroscopic adhesion. Proceedings of the National Academy of
Sciences, 111(9), 3298-3303.

Persson, B.N.J. (2002). Adhesion between an elastic body and a randomly rough
hard surface, Eur. Phys. J. E 8, 385--401

Persson, B. N. J., \& Tosatti, E. (2001). The effect of surface roughness on
the adhesion of elastic solids. The Journal of Chemical Physics, 115(12), 5597-5610

Persson, B. N. J. (2007). Relation between interfacial separation and load: a
general theory of contact mechanics. Physical review letters, 99(12), 125502.

Persson, B. N. J. (2014). On the fractal dimension of rough surfaces.
Tribology Letters, 54(1), 99-106.

Persson, B. N., \& Scaraggi, M. (2014). Theory of adhesion: r\={o}le of
surface roughness. The Journal of chemical physics, 141(12), 124701.

Peressadko, A. G., Hosoda, N., \& Persson, B. N. J. (2005). Influence of
surface roughness on adhesion between elastic bodies. Physical review letters,
95(12), 124301.

\bigskip
\end{document}